# E-COMMERCE IN TURKEY AND SAP INTEGRATED E-COMMERCE SYSTEM

**Ahmet KAYA**
Assoc. Prof. Dr., Tire Kutsan Vocational Training School, Ege University, İzmir, Turkey
Email: ahmet.kaya@ege.edu.tr
Orcid ID: 0000-0002-6105-0787

**Ömer AYDIN** (Correspondence Author)
Dr., Faculty of Economics and Administrative Sciences, Dokuz Eylül University, İzmir, Turkey
Email: omer.aydin@deu.edu.tr
Orcid ID: 0000-0002-7137-4881

## ABSTRACT

E-commerce, it is a kind of shopping by use of internet. E-commerce, very different from the usual shopping concept, it is compatible with today's economic dynamics. E-commerce is becoming an indispensable method with the increase of the internet usage. With the use of E-commerce, there are also a number of advantages for companies. On the other hand, SAP is a pioneer and leader in company resource planning software sector. SAP is very important for large-scale companies. They manage all their processes on SAP and its integration is very important with other related software. In this article, we give a brief information in some important aspects about e-commerce and propose a solution for ERP integration of an e-commerce system.

**Jel Codes:** L81, L86, M00, M31

1. INTRODUCTION

According to the definition of many scientific sources, commerce; defined as any kind of trading activity carried out for gain a profit. Commerce or commercial activity is a phenomenon that has emerged with the existence of humanity and has gone through a series of changes until today. These activities have started in the form of exchange of goods in order to meet the needs of people. In time it has developed due to economic and technological changes. Until today, societies have gone through various stages and it is seen that different commercial procedures are used in each stage.

It is accepted that the societies go through three stages such as agricultural society, industrial society and information society. It is seen that this evolutionary development has been passed from hunter-gatherer / nomadic society to agricultural society, then to industrial society and finally from post-industrial society to information society (Bayraktutan, 2004: 48).

In the information society, which is accepted as today's society, goods and services are introduced to consumers through internet sites and televisions. These promotions make it possible for people to buy more than their needs. Thus, consumption habits of individuals are increased and consuming societies are created. Consumers can easily waste goods produced with great efforts. Thus, production processes are developing and it is possible to produce cheaper. In short, a period begins in which today's economic dynamics are controlled by internet portals. However, the rapidly growing and developing shopping sites offer many campaigns to customers. Opportunities such as shopping points or credit card points increase the willingness of individuals to purchase. The amount of shopping made through these sites is increasing day by day. Shopping via the Internet eliminates the popularity of traditional merchandising. It causes the shops which are located in very good places to suffer from the scarcity of sales. With internet sales, costs are reduced and shipments are made directly from the warehouse to the customer.

According to the household information technology usage survey of Turkey Statistical Institute (TÜİK), the rate of computer and internet usage among individuals aged 16-74 was 75.3% shown in Figure 1. Moreover, eight out of every 10 houses have internet access. 34,1% of individuals with access to the Internet do shopping (E-Commerce) via the Internet. Shopping or electronic commerce via the Internet increased by 4,8% compared to just a year ago. It should be seen that this increase, which can be seen as small, actually represents a huge rate increase. The histogram below represents the computer and internet usage rates for the ages 16-74 years between 2008 and 2019 (Source: TÜİK, 2019).

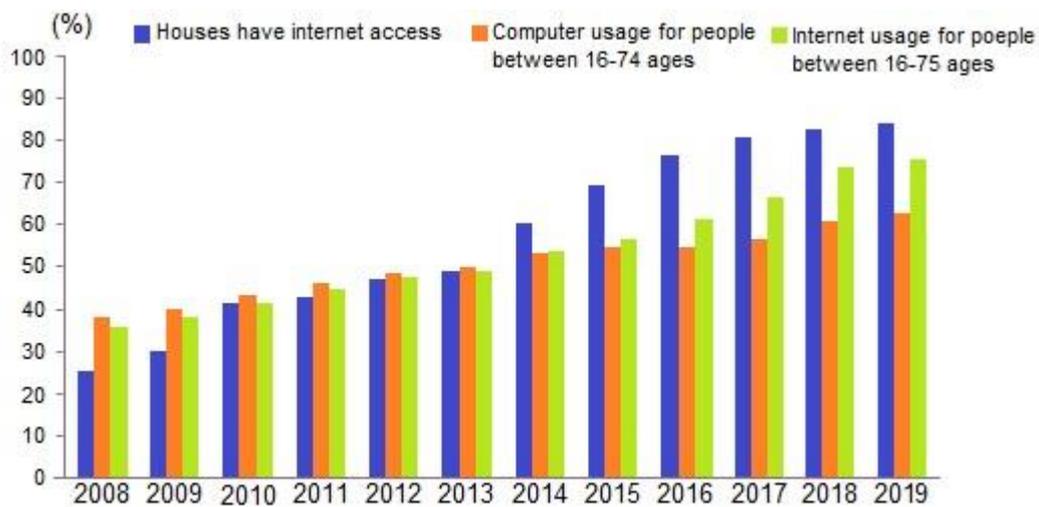

**Figure 1.** Internet and computer usage statistics in Turkey between 2008 and 2019.

The rapid increase in consumption causes the increase in consumption-dependent production compulsory. Thus, production has become continuous. This dynamic process has led to the production being carried out in a system. Production has to be carried out within the framework of scientific procedures. On the cost side of production, serious optimizations have occurred.

However, consumption based on the capitalist system has shifted with technology over time. It is seen that the internet has become an important tool along with some other factors. Visual media, advertisements, social platforms and sharing networks are the determinants of the increase in consumption (Halis, 2012).

The development and diffusion of e-commerce makes a significant contribution to the development of the logistics transportation sector. E-Commerce contributes to the development of a commercial activity area called E-logistics and the services provided in this area become more serious and institutional (Gülenç & Karagöz, 2008: 78).

Another contribution of e-commerce to national and world economy is to register all trade. Informal economy, which is one of the important problems of developed countries, is completely eliminated and considerable improvements in tax revenues are achieved.

According to the results of the global research company, e-Marketer, the volume of e-commerce is estimated to reach 1 trillion 700 billion dollars in 2015, 2 trillion dollars in 2016, and 2 trillion 500 billion dollars in 2018 (Sabah, 2015).

As we have seen, the share of trade held by the Internet in Turkey and in the world is increasing every day. As a developing country Turkey is showing remarkable increase in the level of 4% a year of commerce.

These huge changes in the field of trade are called the new economy. New economic relations based on information technologies can be defined as new business opportunities and the reshaping of existing business areas using new communication environments (Baily and Lawrence, 2001: 8).

Turkey has made significant strides in recent years to get a position in the process of the new economy. Turkey's first objective is using and generalizing of information and communication technologies within the country which is one of the basic conditions of the EU membership (Barışık and Yirmibeşcik, 2006).

The economic effects of electronic commerce lead to an acceleration of the transition process from traditional trade understanding to the new economy in the world (Atınok, Sugözü and Çetinkaya, 2007).

## 2. E-COMMERCE

There are many different definitions and explanations about E-Commerce. Sometimes these explanations are limited to certain areas of trade. These limitations do not actually conform to the nature of E-Commerce. This is because the definition made by the US-based National Telecommunications and Information Administration is considered to be much more realistic and accurate. According to this agency, E-Commerce is defined as usage of all kinds of electronic technologies for commercial activities in desired form (Isler, 2008: 278). From this perspective, the concept of E-Commerce includes all kinds of shopping activities via the Internet and or any other technological environments. Some of the commercial activities carried out via the Internet are as follows: TV broadband subscription services, mobile phone subscriptions, prepaid phone cards, stock and exchange transactions, banking activities, insurance, hotel accommodation reservations, travel tickets, car rental, sports and cultural activities, education and training, examination services and all other goods.

E-Commerce includes a number of innovations that make life easier for poeple today. For example, banks can make billions of routine banking transactions only through internet banking. It is impossible to make such a large number of transactions through bank branches. There are at least 3 or 5 bills that each family pays with automatic instructions. This has made it very easy for people to make their banking transactions, so people of all ages prefer to do so.

In addition, all operations of our schools, where education and training activities are carried out, are carried out via the internet in connection with the financial centers. Mission undertaken by the ÖSYM, Turkey's central exam center, is so developed and grown by the internet. Otherwise, it would not be possible to make exams for 20 million people in a year and announce the results within a very short period of time. On the other words, if banking transactions are accepted as a commercial activity for banks, it is not possible to accept other transactions performed under these transactions independently from E-Commerce. Given this situation, it is not possible to define the limits of E-Commerce.

It is necessary to mention the definition made by the Organization for Economic Cooperation and Development (OECD), which is one of the important institutions in the field of E-Commerce. According to OECD, E-Commerce are called transactions that are based on the transmission of digitized data such as written text, voice, and images through open or closed networks with the participation of individuals and institutions (Kalaycı, 2008: 140).

E-Commerce provides a number of advantages for the companies that carry out this activity. These advantages are largely economic advantages. It gives so many advantages such as reducing or eliminating the store rent expenses, personnel expenses, water, electricity, natural gas and a number of other expenses so the input costs of the products can be reduced. Moreover, there is no need to worry about the store location and customer satisfaction from the store staff. The functionality of the designed shopping website and the proper display of the products is sufficient. There are also a number of advantages for the purchaser of the product: Product can be examined more detailed in that web sites and compared with many products in a few minutes and purchased cheaper than classical stores.

If we considering the opportunities offered by the internet and the use of the web sites we can see that E-Commerce enables competition in market. These considerations are due to the elimination of costs through internet technology, the real-time acquisition of market data, and the easy access of accurate information to consumers (Ellison and Ellison, 2005).

The World Trade Organization (WTO) defines E-Commerce as follows: E-Commerce defines the production, advertising, sales and distribution of goods and services through telecommunication networks. According to this definition, which is considered to be very ambitious and different, the trade process has three basic stages (WTO, 1999: 2).

1. Research stage of producers and consumers or buyers and sellers,
2. Making the order and payment,
3. The stage of delivery of goods to the buyer (Kalayci, 2008: 141).

If we consider E-commerce in the framework of the consumer E-Commerce covers searching, finding, ordering, making payments and monitoring the process. On the other hand, if we consider it in the framework of the company, it covers the processes of product promotion, taking orders, collecting fees, sending the products and providing service. (Fidan and Albeni, 2014: 288).

**2.1. Types of E-Commerce**

E-Commerce is a concept that is changing and expanding its borders in electronic world. This trade, which may be national and international, can be realized in many different ways. Accordingly, E-Commerce types can be found in 6 categories.

1. Business-to-business E-Commerce (B2B),

2. E-Commerce between businesses and consumers (B2C-Business to Consumer),

3. E-Commerce between enterprises and public administration (B2G-Business to Government),

4. E-Commerce between consumers and public administrations (C2G-Consumer to Government),

5. E-Commerce among consumers (C2C-Consumer to Consumer),

6. E-Commerce between States (G2G-Government to Government) (Kalayci, 2008: 142).

The most widely used E-Commerce types are; business to business (B2B), business to consumer (B2C), consumer to consumer (C2C) ((Laudon and Traver, 2010: 17).

**2.2 E-Commerce Strategies**

E-Commerce has become a strategic trading tool by providing a set of strategic advantages for businesses (Isler, 2008: 284). The main factor that strengthens this strategy is that people prefer to meet in virtual environments instead of meeting face to face. In this context, strategic use is realized within the framework of three elements (Sadowski, 2002: 78).

These three elements;

1. Communication requirements of people,

2. Competition policies of enterprises,

3. Economic and social life needs.

It is unthinkable not to meet these issues which are very important for businesses and people. People who are in danger-free internet environments are shaping their shopping habits as they do not incur additional costs. In addition to the stores in shopping centers, opening virtual stores becomes a strategy for the companies. The shopping behavior of the customers has become a matter of quality and price balance. Some of the cost independent products are offered to the market at a cheaper price with e-commerce, thus making them the consumer's choice.

Also known as virtual commerce, E-Commerce is a trade area in which the smallest intellectual can realize his thoughts and actions without capital and investment thanks to information technologies (Crimea, 2001: 31). Due to this nature, it is accepted that E-Commerce is a post-modern trade approach (Brown, 1995).

When these qualities of e-commerce are taken into consideration, it is necessary to acknowledge that an easy and effective tool of commerce is encountered and this is an important, valuable and strategic tool.

**2.3 Advantages and Disadvantages of E-Commerce**

Electronic commerce, which is considered to be one of the most important trading and marketing tools of today, provides many advantages and new business opportunities to visionary firms, and enables consumers to purchase goods and services cheaper and easier. As the Internet and communication opportunities develop, these benefits are constantly evolving and increasing. The advantage of electronic commerce is mainly due to lowering costs related to commercial activities, eliminating geographical limitations and increasing transaction speed (Güleş, Bülbül and Çelebi, 2008: 468).

The advantages of E-Commerce are as follows:

Advantages

• Transactions are fast and instant,

• There are no time barriers to trade. 7/24 continuous operation,

• Profit and cost optimization,

• It is easier to win customers with attractive price opportunities,

• It is possible to find new business partners. In this way, it is possible to establish international partnerships and expand the trade volume,

• Becoming a brand is easier for businesses,

• Supply of raw materials and products is easier and cheaper,

• Trade operations are always recorded, guarantees are real,

• Tax loss and informality disappear,

213

• Since the supply of goods and services is fast, production and marketing are also fast.

Disadvantages

• It is not possible to convince the customer with the right communication as in the real trade.

• The ideal trade laws required by virtual commerce have not yet been enacted,

• It is possible to be exposed to internet irregularities,

• E-Commerce cannot develop in those who do not use the Internet,

• There is no synergy provided by the shopping and market environment,

• It is not possible to see the product and be informed about the quality of the product,

• Lack of social trade relations,

• Traditionalism is lost.

**2.4 Risks in E-commerce**

While E-Commerce provides a number of convenience and advantages in the field of commerce, it naturally inherits certain risks. The most important risk of E-Commerce is, of course, security. In addition to security concerns, the fact that shopping via the internet is not adequately secured in the trade laws, virtual sites created by malicious persons, copying credit card information to third parties, not paying the products to the address, products being different from the defined products, product returns and not to be committed to commitments, marketing of used products and similar risks. Based on the existence of all these risks, it is preferable to respect the products of commercial organizations that are confident of their corporate identity and to shop on their sites. Customer reviews about the company and products must be read before shopping through the company portals. It should be noted that customer reviews in particular are a summary of some important experiences.

**3.     STORE DYNAMICS**

E-Commerce is committed to transforming the foundations of commercial and non-commercial life. (Kalayci, 2008: 140). Retailing in Turkey continues its activities in shopping areas such as traditional trade stores. As a developing economy, competition and product quality are not at the highest levels. In retailing, products of the same quality are marketed with different pricing policies. Therefore, there is a lack of trust due to insufficient protection of consumers' rights. The retail sector, where competition is not at a high level, faces a series of trust problems. The desire of consumers to access the perfect market is considered as a factor enabling E-Commerce to be accepted.

There are a number of precautions that retailing needs to take in order to avoid this negativity and at least to protect its customer portfolio. For example, the positive and negative effects of the store atmosphere on the employees and customers are within the scope of the research. (See Oğuz and Gürdal, 2017). In addition, even the level of education and age of marketing staff is considered a dynamic in retailing (Yüksekbilgili and Akduman, 2015). All these precautions are necessary and important, but it should be noted that these innovations are cost-increasing factors.

**3.1 Risks in E-Commerce**

As a matter of fact, this form of trade was accepted as ideal in the first years when E-Commerce entered human life (Bakos, 1998: 35). However, in the following years, this view was taken away and it was accepted that E-Commerce contains some faults (Fidan & Albeni, 2014: 289). Despite all its shortcomings, E-Commerce has become a preferable form of trade in terms of quickly finding and comparing multiple options, comparing products, analyzing prices, and accessing product and store reviews.

Corporate companies have implemented E-Commerce activities together with retailing. Thus, they made their commercial activities alternative. For these companies, shopping activities have stabilized and developed continuously. It is not difficult to predict that there will be a decrease in the sales of the companies which continue their activities by doing only retailing.

In this respect, retailing is traditional but not up to date. Therefore, there are difficulties and cost increases in the supply of raw materials and products. A survey showed that the cost of the product sold in the store increased by 33% until it came to the shelf. (Civelek and Sozer, 3003: 120). This cost increase is very important and is an additional cost that reduces competition in retail.

E-Commerce is a form of commerce that governments want and tries to implement. This is because all commercial transactions are recorded and there is no tax loss. However, it is possible to identify a number of unregistered transactions in shopping from a neighborhood market. It is even possible to come across practices that reduce prices of products and services if a receipt or invoice is not requested in some cases.

As a developing country, Turkey, has to increase tax revenue and record all trading activities. E-Commerce has the dynamics to provide these operations. This is because the trade is made in the form of wire transfers or EFT (Electronic Funds Transfer) through credit cards or banks.

## 3.2 Advantages and Disadvantages of Retailing

Retailing has a number of advantages that contribute to the development of trade and market phenomena. The basic dynamics that keep the merchandising alive are being traditional, providing a physical environment for the customer, direct communication between the seller and the customer, having a shopping ambience, supporting sales with the advice of the staffs, and having an idea by seeing the goods to be purchased. However, the business of retailing becomes more difficult as customers' alternatives and sources of information increase and diversify (Oğuz & Gürdal, 2017: 42). It is not possible to mention the dynamism of retailing, physical environment and other dynamics in E-Commerce. The advantages and disadvantages identified for retailing are listed as follows:

### 3.2.1 Advantages of retailing

• Being able to benefit from vendor consultancy,

• Possibility of tasting, testing or seeing according to the nature of the marketed product,

• Bargaining in price,

 • The habit of being at a specific address,

• One-to-one comparison with other brand products.

### 3.2.2 Disadvantages of retailing

• Having a fixed location,

• Loss of customers due to the negative synergy of vendors,

• The need for continuous renewal of shop design,

• Problems caused by rapid staff change,

• Problems caused by communication problems between employees,

• The intensity of competition and loss of customers resulting from the presence in the alternative stores platform,

• The necessity of movement of goods due to fashion and similar reasons.

## 4. E-COMMERCE WITH NUMBERS

According to TÜİK data for 2019, some statistical informations about E-Commerce are as follows (TÜİK,2019):

In 2019, the rate of poeple who did not make e-commerce transactions even though they used the internet was 65.9%. The reasons for not shopping online are as follows:

**Table 1:** Reasons for not use E-Commerce

| Reasons | Rate |
|---|---|
| The desire to see the product and the habit of shopping in the store | 81,2 |
| Lack of knowledge and skills | 21,0 |
| Problem of delivery of goods | 20,3 |
| Privacy and security concerns without paying | 44,9 |
| Mistrust in returning the product | 27,0 |
| Not have a credit card | 16,2 |
| Other | 4,0 |

According to the data of 2019, the proportional distribution of the types of goods and services purchased by customers using the Internet through E-Commerce in the last 12 months can be shown in Table 2.

**Table 2:** 2019 Types of Goods and Services in E-Commerce for personal use in the last 12 months

| TYPES OF GOODS AND SERVICES | RATE |
|---|---|
| Daily requirements and food | 27,4 |
| Household appliances, furniture and other goods | 26,9 |
| Medicine | 4,1 |
| Clothes and sports equipment | 67,2 |
| Computer and hardware | 12,1 |
| Mobile phone and other electronic devices | 20,3 |
| Telecommunications - services | 15,3 |
| Holiday accommodation | 14,8 |
| Travel ticket and car rental | 31,7 |
| Cinema, theatre, concerts, etc. services | 18 |
| Movies, music | 9 |
| Books, magazines and newspapers | 20,2 |
| E-learning tools | 3,3 |
| Games and other software | 6,6 |

The distribution of individuals having problems in purchasing and placing orders during the E-Commerce activity in 2017 is shown in Table 3.

**Table 3:** Problems in E-Commerce

| Problem | Rate |
|---|---|
| Problem with order or payment | 20,1 |
| Failure to obtain information about warranty terms and other | 13,6 |
| Late delivery of goods | 46,5 |
| Unexpected transaction costs issue | 9,8 |
| Wrong or damaged product delivery | 49,1 |
| Fraud | 15 |
| Problems encountered in complaints and compensation | 19,5 |
| The problem of not receiving goods from foreign sites | 7,3 |
| Other | 1,8 |

In this context, new companies are organizing their strategies with the principle of "customer-oriented" work. The quality of the companies in terms of service and customer satisfaction is a showcase. In addition, in line with the demands of today's consumers, companies focus on product quality assurance, short lead times and after-sales service.

The primary objective for the companies is to maximize profit. In order to increase the profitability level, it should increase inventory turnover and cash flow control, reduce the changing costs and make its appearance on the financial front more effective. Companies can increase their competitiveness by considering customer-oriented logic and firm's internal balance. Considering today's selective customer portfolio, many companies known as global brands have been able to success about these criterias. Other companies have to keep up with this situation and control their resources in the best way.

ERP software approaches the company with a method based on the production process. ERP integrates all functions in a certain order considering the company's objectives. It is a modern software solution that can be run in many auxiliary systems in order to meet the information and data needs. In this context, ERP system can be seen as a main software system for production, finance, retailing, e-commerce and etc. It has become the name of the information systems infrastructure that supports all corporate processes.

The ERP system, which stands for "Enterprise Resource Planning", has emerged as a result of eliminating the deficiencies in the Material Requirement Planning (MRP) and Manufacturing Resource Planning (MRP-II) systems (Fui-Hoon, F., 2002).

## 6. SAP

SAP AG is a software company that provides specialized software solutions to companies that are not large-scale in the field of company operations. In addition, SAP AG software enables both small and medium-sized companies operating commercially to operate in an integrated manner, both individually and interacting with each other. Furthermore, in 1972, with the initiative of many analysts, they established the System Analysis and Program Development in Mannheim (Germany). Today, it is possible to say that there are offices in almost every part of the world.

The objective of SAP is to provide support to the senior staff of the company in optimal business solutions in all sectors and especially in large-scale companies. It is also to meet all needs as enterprise resource planning software.

We can examine the development of SAP software, which has developed over the years and made it suitable for today's conditions, as detailed below.

On April 1, 1972, five of IBM's founders founded SAP in Mannheim, Germany. Their goal was to produce and sell standard enterprise software that could integrate all commercial and administration-based applications. In 1977, SAP software developers moved from Mannheim to Walldorf. In 1978 SAP, together with the R/2 software system, took a solid step in the development process. In 1982, in the 10th anniversary of its founding, SAP proved its success by increasing its sales by 48% and exceeding the level of 24 million Deutsche Marks. After 1982, 236 companies, mainly Germany, Austria and Switzerland, continued their activities with SAP software. SAP International was founded in 1984 in Switzerland. In 1985 sales reached 100 million Deutsche Marks. In 1987, the use of SAP application reached 850 companies. SAP released a new sub-software (R/3 system) in 1992. In 1995, Microsoft joined SAP as a R/3 customer. In 1997, different solutions including Customer Relationship Management (CRM) and Supply Chain Management (SCM) were added to the scope. In 1999, Enjoy SAP application aimed to make the learning and use of the software easier. With mySAP.com, it became the leading e-commerce software of the 2000s with its way of working with existing ERP solutions in e-commerce and web technologies. The mySAP strategy later found finally as SAP NetWeaver. SAP NetWeaver has become a highly flexible and compatible application for complete support of production processes. In December 2011, SAP acquired Success Factors, the largest cloud developer (SAP History, 2019).

**6.1 SAP Integrated E-commerce**

The proposed solution in this article consists of 3 basic layers. All detailed can be seen in Figure 2. These are Main system layer, intermediate database layer and e-commerce system layer. In these three layers, three separate database systems are used. An independent database has been used in the e-commerce system layer due to the difficulties in synchronization between SAP and e-commerce software. The intermediate layer, where the information is temporarily stored prior to importation into SAP ERP, includes the Oracle database. This layer can also include VPN server and Firewall. Another database in which SAP ERP system's own data is kept is Oracle. In this solution, a design can be created by not using the middle layer, which is the second layer, by directly transferring the data received from the e-commerce system to the database used by SAP ERP. However, as a result of the investigations, it was determined that this would cause some problems and an independent database system was added to the design.

The reasons for using the independent intermediate database management system;

• Because the ERP system is alive and in use by the whole company, uncontrolled connection of the client to the system may cause performance problems.

• Since there is no data that can be processed directly from the clients to the system, these data are taken into the intermediate database and processed at any time and taken over the ERP to create a more accurate design.

• Creating records directly in the system means automatic discarding of accounting records. This can lead to incorrect and uncontrolled financial data.

• Even if the SAP ERP system is taken as an open structure, even if all precautions are taken, putting a layer together will reduce the security problems.

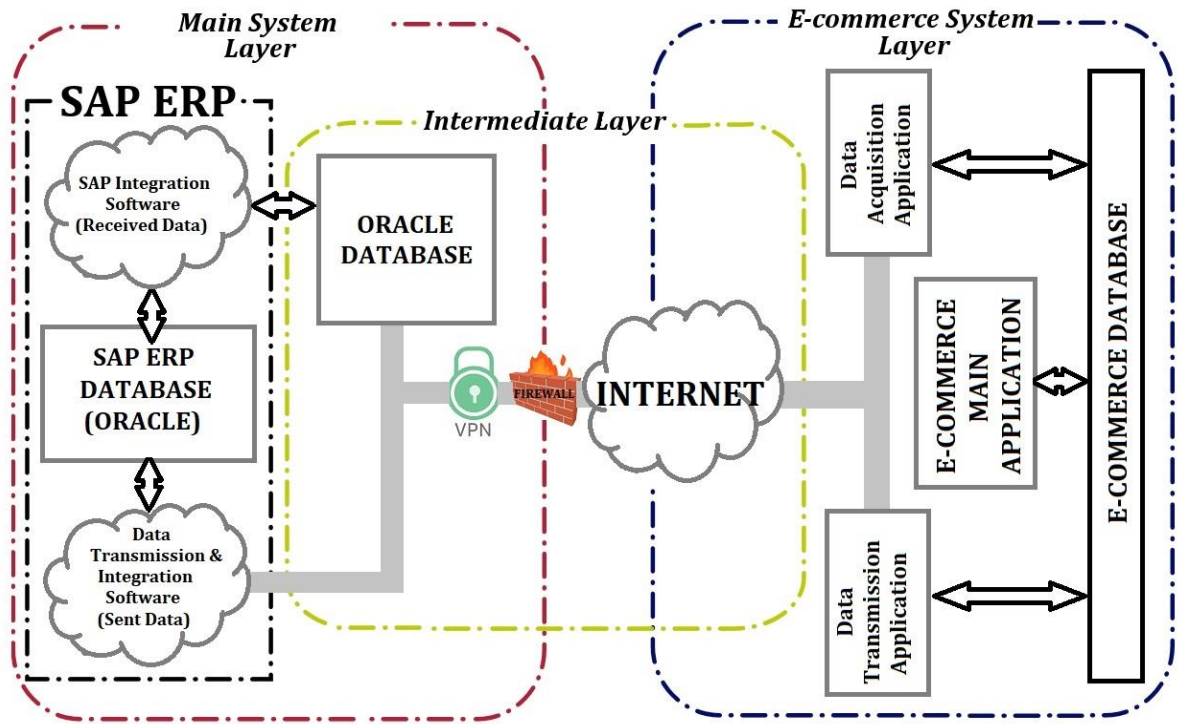

**Figure 2.** System technical view

### 6.1.1 Main System Layer

SAP ERP software supports a large number of database management systems. Software can be installed by selecting one of these databases. SAP supports the databases listed below:

- Oracle
- MS SQL Server
- IBM DB2 for Linux, UNIX, and Windows
- SAP live Cache Technology
- SAP MaxDB
- IBM DB2 Universal Database for z/OS
- IBM DB2 Universal Database for iSeries (SAP, 2019)

In our application, Oracle database was used in SAP installation. SAP has restricted direct access between the database and the user. SAP has two layers between user and database. The first layer is the presentation layer. Users can access this layer. The operations performed here are transmitted to the middle layer which is a sub-layer. The middle layer interprets incoming transaction codes and passes them to the database.

**6.1.2 Intermediate Layer**

In this study, records created as a result of e-commerce transactions are first kept on the Oracle database which we call intermediate layer and then transferred to SAP. As a result of this transfer, orders, deliveries, invoices and accounting documents are generated for sales on SAP. Since these records are standard SAP processes, all data are generated in SAP standard tables.The SAP Java Connector (SAP JCo) is used to connect to SAP. SAP JCo is a development library that enables a Java application to communicate with on-premise SAP systems via SAP's RFC protocol. It combines an easy-to-use API with unprecedented flexibility and performance (SAP Java Connector, 2019).

A copy of the e-commerce system database is created in the intermediate layer. Thus, the data in the e-commerce system will be provided to the central server. This database will basically keep the sales data and product information that are created in e-commerce system and transferred to ERP system.

**6.1.3 E-commerce Layer**

This layer enables the e-commerce database to be transferred to central system. Likewise, it transfers product information, prices and other necessary data from the central system layer. In this layer, there are data reception, data transmission programs, e-commerce software and database.

In this layer, the changes made on SAP ERP are reflected in the e-commerce system. For example, product updates, deletions and new product registration are transferred to e-commerce system in this layer.

Unlike the upload application, this layer retrieves data by connecting directly to a function on SAP ERP instead of the using the intermediate database. SAP ERP allows data exchange with several different methods, as well as externally accessible functions. For this operation, a user with remote access and required permissions must be created. Additional packages are required for the connection, which include classes for this user and some connection specific to the programming language used (this connection class is SapJCo because we use Java). All these adjustments have been made on SAP ERP. If the data exchange between the application and the SAP ERP function is multiple, the data structure is performed in the form of series. This data structure will be taken as a table on the Java side and the data in it can be accessed.

**7. CONCLUSION**

E-Commerce is an increasingly used form of purchasing goods and services in Turkey as in the whole world. It is developing and becoming widespread due to the density of internet usage. Considering the close relationship between e-commerce and information technologies, it is accepted that it is implemented more effectively and accurately in the USA and European countries. Electronic commerce seems to be safer in countries that place trade laws and Internet-connected crimes on legal platforms. It is necessary to accept that the biggest risk to E-Commerce is security-based problems.

According to TÜİK, internet usage rates in Turkey is increasing rapidly. The rate of internet users between the ages of 16-74 has reached 75.3%. 34.1% of these users carry out trading activities through E-Commerce. According to these results, the proportion of individuals between 16-74 who trade in Turkey is 25.6%.

While these changes are taking place on the e-commerce side, a number of problems are experienced in the retailing. In retailing, store rent, dues, water and electricity payments, personnel expenditures and other expenses are factors that increase the price of the goods offered for sale in certain proportions. In this case, customers who are intertwined with technological platforms may have the opportunity to purchase the goods of the same companies through the internet at a much cheaper price. However, it is understood that 81.2% of the consumers who do not prefer to shop through E-Commerce act with the desire and habit of seeing the goods in the store. This is followed by security concerns with a rate of 44.9%. It is understood that customers do not prefer to shop with E-Commerce platforms since they think that he will have difficulty in returning the product at a rate of 27.0%. Other causes can be traced in Table 1.

The types of goods purchased by customers shopping through E-Commerce were as follows: 67.2% of the trade was realized as the purchase and sale of sports equipment. This type of shopping is followed by travel tickets and car rental with 31.7%, Daily requirements and food with %27.4, Household appliances, furniture and other goods with 26.9% and book magazines and stationery products with 20.2%. See Table 2 for other rates. Based on these data, it is necessary to see that sporting goods, white goods and food products take the lead and these sectors are essentially merchandising products.

While E-Commerce incorporates a series of developments, innovations and ease, it is necessary to know that it includes some problems. According to TÜİK data, the main problem encountered in E-Commerce was determined as the problem encountered in the Wrong or damaged product delivery with a rate of 49.1%. Late delivery of goods is in the second place with 46.5% and order and payment problems are in the third place with 20.1%. After that, the problems encountered in the complaint and compensation, fraud and other problems listed in Table 3.

It should be noted that while the sales increase through the internet, retailing is affected but does not lose its attractiveness, because it is important for the conservative societies that adhere to the traditions. In order to increase the diminished customer interest, retailing is structured within the shopping centers and a series of attraction-enhancing measures are implemented. These measures trigger customer interest and increase the price of the goods sold. This result naturally leads to a situation that leads the customer to E-Commerce again.

The ERP sector has been dominating the SAP ERP software industry for many years. The SAP Enterprise Resource Planning (ERP) application is designed to effectively support the core functions and operations of business processes and meet industry-specific requirements. However, it is not possible for ERP SAP software to fully respond to the needs of the developing and spreading sectors. For some special cases, SAP ERP had to develop software that can work integrated with it and the necessary technological infrastructure was provided for this situation.

This article describes how a company using SAP ERP and doing electronic commerce can integrate two software. This type of developed integration softwares run as a part of ERP and enable enterprises to use the data integratedly without losing the advantages of ERP software. The proposed integration solution provides a secure, manageable, scalable structure.

## REFERENCES

Baily, M., Lawrence, R.Z. (2001). Do We Have a New E-conomy, *National Bureau of Economic Research,* Cambridge, MA, Working Paper, No:8423.

Bakos, J.Y. (1998). The Emerging Role of Electronic Marketplaces on the Internet,

*Communications of the ACM*, 41(8): 35-42

Barışık, S., Yirmibeşcik, O. (2006). Türkiye'de Yeni Ekonominin Oluşum Sürecini Hızlandırmaya Yönelik Uyum Çabaları, *ZKU, Sosyal Bilimler Dergisi*, Cilt:2, Sayı:4, Sayfa:39-62.

Bayraktutan, Y. (2004). Bilgi İktisadi Geliştirme Evreleri ve Maldan Sanala Paranın Evrimi,

*Türkiye Günlüğü Dergisi*, Sayı:78.

Brown, S. (1993). Post Modern Marketing, *European Journal of Marketing*, Cilt No:4.

Civelek, M. Sözer, E. (2003). İnternet Ticareti: Yeni Ekososyal Sistem ve Ticaret Noktaları,

*Beta Yayınları*, İstanbul.

Ellison, G. Ve Ellison, S. F. (2005). Lessons about Markets from the Internet, *The journal of Economic Perspectives*, 19(2): 175-193.

Fidan, H. Ve Albeni, M. (2014). Asimetrik Bilginin E-Ticaret Üzerindeki Etkileri: Tüketicinin Güven Eğitimleri Üzerine Bir Araştırma, *Süleyman Demirel Üniversitesi İktisadi ve İdari Bilimler Fakültesi Dergisi.* Cilt:19, Sayı:2, Sayfa287-298.

Gülenç, F. Karagöz, B. (2008). E-Lojistik ve Türkiye'de E-Lojistik Uygulamaları, *Kocaeli Üniversitesi Sosyal Bilimler Enstitüsü Dergisi* (15) 2008 / 1: 73-91.

Güleş, H.K., Bülbül, H. ve Çelebi, A. (2008). Küçük ve Orta Ölçekli Sanayi İşletmelerinde Elektronik Ticaret Uygulamaları, *Sosyal Bilimler Enstitüsü Dergisi*, Konya.

Halis, B. (2012). Tüketimin Değişen Yüzü: Elektronik Ticaret Uygulamaları ve Sosyal Paylaşım Ağlarının Rolü, *Tarih, Kültür ve Sanat Araştırmaları Dergisi*, ISSN:2147- 0626, Vol:1, No:4, 2012.

İşler, D.B. (2008). Rekabetçi Avantaj Yaratma Çerçevesinde Kobilerde E-Ticaret ve E-Ticaretin Stratejik Kullanımı, *Süleyman Demirel Üniversitesi İktisadi ve İdari Bilimler Fakültesi Dergisi*, Vol:13, No:13, Sayfa: 277-291

Kalaycı, C. (2008). Elektronik Ticaret ve Kobilere Etkileri, *Uluslararası İktisadi ve İdari İncelemeler Dergisi*, Yıl:1, Cilt:1, Sayı:1, ISSN:1307-9832.

Kırım, A. (2001). Strateji be birebir pazarlama CRM, *Sistem Yayınları*: 266 İstanbul.  Laudon K. ve Traver, C.G. (2010). Introduction to E-commerce, Prentice Hall, New Jersey.

Oğuz, V.G. ve Gürdal, S. (2017). Parekende Sektöründe Mağaza Atmosferinin Satış Personeli Performansı Üzerine Etkileri: Hazır Giyim Sektöründe Bir Araştırma, *International Journal of Academic Value Studies, (Javstıdies)*, ISSN:4149-8598, Vol:3, Issue:8, pp:41-54.

Sabah, (2015). İnternetten Alışveriş Raporu, *Sabah Gazetesi, 30.04.2015 tarihli ekonomi raporu*.

Sadowski, B. M., Maitland, C., Van Dongen, J. (2002). Strategic Use of the Internet By Small

and Medium-Sized Companies: An Exploratory Study. *Information Economics and Policy*.

TÜİK, (2017). Hanehalkı Bilişim Teknolojileri Kullanım Araştırması, *Haber Bülteni*, 27 Ağustos 2019, Sayı: 30574.

Yüksekbilgili, Z., Akduman, G. (2015). Parekende Mağazacılık Sektöründe Satış Personelinin Demoğrafik Özellikleri ve Personel Memnuniyet İlişkisi, *Elektronik Sosyal Bilimler Dergisi,* Kış-2015, Cilt:14, Sayı: 52, Sayfa:86-99.

SAP History (2019). SAP: A 47-year history of success, https://www.sap.com/corporate/en/company/history.html, 30.09.2019

Fui-Hoon, F., 2002, Enterprise Resource Planning Solutions and Management, Travers, J., IRM Press, USA.

SAP , Supported Database Systems (2019). https://help.sap.com/viewer/d77277f42c0b469db8794645abd954ea/8.0/en-US/64f4ef081ed74836b570b56d7bcb4527.html, 30.09.2019

SAP Java Connector (2019), https://support.sap.com/en/product/connectors/jco.html, 30.09.2019